\documentclass[aps,prl,reprint,twocolumn,superscriptaddress]{revtex4-2}
\usepackage{xcolor}
\usepackage[normalem]{ulem}
\usepackage{soul}

\usepackage{amsmath}
\usepackage{amssymb}
\usepackage[paper=a4paper,left=25mm,right=25mm,top=25mm,bottom=25mm]{geometry}
\usepackage{mathtools}
\usepackage{dsfont}
\usepackage{hyperref} 
\usepackage{cleveref}
\crefname{figure}{Fig.}{Fig.}
\crefname{equation}{Eq.}{Eq.}
\usepackage{siunitx}
\usepackage{subcaption}
\usepackage{graphicx} 
\usepackage[sort&compress]{natbib}
\graphicspath{{Graphen}}
\usepackage{float}
\usepackage{orcidlink}
\begin{document}
	\title{Universality Emerging in a Universality: Derivation
	of the Ericson Transition in Stochastic Quantum Scattering
	and Experimental Validation}
\author{Simon
	K\"ohnes,\orcidlink{0009-0001-2993-5315}}\email{simon.koehnes@uni-due.de}\affiliation{Fakult\"at
	f\"ur Physik, Universit\"at Duisburg-Essen, Duisburg,
	Germany} \author{Jiongning Che, \orcidlink{0000-0001-8512-8991}} \altaffiliation[Present Address:
]{Yangtze Delta Region Institute, University of Electronic
	Science and Technology of China, Huzhou, China}
\affiliation{Lanzhou Center for Theoretical Physics, Lanzhou
	University, Lanzhou, Gansu, China} \author{Barbara Dietz,\orcidlink{0000-0002-8251-6531}}
\email{bdietzp@gmail.com} 
\altaffiliation[Present Address:
]{Max-Planck Institute for the Physics of Complex Systems and TU Dresden, Dresden, Germany}
\affiliation{Lanzhou Center for Theoretical Physics, Lanzhou
	University, Lanzhou, Gansu, China}
\affiliation{Center for Theoretical Physics of Complex Systems, 
	Institute for Basic Science, Daejeon, Korea}
 \author{Thomas Guhr,\orcidlink{0000-0002-0927-6324}}
\email{thomas.guhr@uni-due.de} \affiliation{Fakult\"at f\"ur
	Physik, Universit\"at Duisburg-Essen, Duisburg, Germany}
\date{\today}
	
	\begin{abstract}
		At lower energies, the resonances in scattering
                experiments are often isolated. In quantum chaotic
                many--body, disordered or generically stochastic
                systems, the resonances overlap at larger
                energies. Eventually, the Ericson regime is reached in
                which the cross section behaves like a random
		function. The scattering--matrix elements then follow a
                universal Gaussian distribution.  For more than sixty
                years, the emergence of this robust additional
                universal behavior on top of the universal system
                stochasticity has awaited a concise analytical
                treatment. We derive the transition to the Ericson
                regime in the universal Heidelberg approach and prove
                the universal Gaussian distribution by a proper
                asymptotic expansion. We also obtain explicit formulae
                for the moments and for the cross--section distribution. We compare with
                microwave experiments and numerical simulations.
	\end{abstract}
	\maketitle

        \textit{Introduction} --- Scattering experiments are
        ubiquitous in quantum and classical physics.  Numerous
        examples for, in a broad sense, chaotic or, more generally,
        stochastic dynamics are found in nuclear
        \cite{BRINK196377, VONBRENTANO1964,
          Porter1965,HARNEY199035,Main1992, Feshbach1993,
          Concepts,Weidenmüller, Weidenmüller2,Frisch2014,
          Kawano2015,SmilanskyIrregular}, atomic and molecular physics
        \cite{LOMBARDI1993, GREMAUD1993, DUPRET1995, SCHINKE1995,
          Scott1996,
          PhysRevLett.95.194101,PhysRevLett.95.263601,Mayle2013},
        quantum graphs \cite{Quantengraphen, QuantumChaos,
          QuantumChaos2,Ławniczak,Celardo2007,Schäfer2003}, microwave
        cavities and networks \cite{
          PhysRevLett.65.3072,PhysRevLett.74.62,PhysRevLett.94.144101,
          PhysRevE.74.036213, PhysRevE.81.036205,
          Dietz2010,PhysRevLett.100.254101,Hul2005,Lawniczak2008,Lawniczak2020,Chen2021}
        and classical wave systems \cite{Weaver,PhysRevLett.75.1546,
          GROS2014664}. The theory of electronic transport
        \cite{RevModPhys.69.731,PhysRevB.35.1039,RevModPhys.72.895,
          PhysRevLett.69.1584, MIRLIN1994325,
          Folk1996,BringiChandrasekar2001, Yeh2012} is closely
        related.  The field of stochastic quantum scattering
        originated in nuclear physics and was pioneered by Ericson in
        the 1960s \cite{Ericson1, Ericson2, Ericson3}. He showed that
        the cross section versus energy behaves like a random function
        when the resonances strongly overlap. The Ericson regime was
        first identified in nuclei \cite{Ericson1, Ericson2,
          Ericson3}, see a review in \cite{Weidenmüller2}, and
	subsequently in numerous systems, \textit{e.g.}, \cite{SmilanskyIrregular}, in electronic
        transport \cite{HAW90,SICZB09}, in atoms
        \cite{Bluemel96,GD97,PhysRevLett.95.194101,PhysRevLett.95.263601,EM09,EM10,EKM11},
        in ultracold atomic-molecular systems~\cite{MRB12}, in
        Bose--Einstein condensates \cite{HHCKB12} and quantum graphs
        \cite{ZM13}. In the Ericson regime, a universal Gaussian
	distribution of the scattering matrix elements emerges. 
       Up to now, it was merely
        phenomenologically understood, here we provide the full
        explanation.
	
	\textit{Scattering process} --- The scattering process is
	encoded in the complex and unitary scattering matrix $S$
        \cite{Mahaux, AGASSI1975,VWZ}. If $M$ channels are connected
        to the interaction zone described by a Hermitean Hamilton
	operator, the elements of the $S$ matrix are
        $S_{ab} = \delta_{ab} - 2\pi i W_a^\dagger G W_b$, where
	\begin{align}
		G &=\left( E\mathds{1}_N - H +i\pi\sum\limits_{c=1}^{M} W_c W_c^\dagger\right)^{-1}
	\end{align}
	is the $N\times N$ matrix resolvent at energy $E$ in a matrix
        notation. The limit $N\to \infty$ has to be taken, $H$ is the
        Hamilton matrix and the $N$--component vectors
        $W_c,\ c=1,...,M$ describe the coupling between bound states
        in the interaction zone and the channels. One may assume their
        orthogonality.  Apart from kinematic factors, the cross
        sections are given by $\sigma_{ab} = |S_{ab}|^2$.

        \textit{Heidelberg approach} --- In a generic chaotic, complex
        or stochastic setting, $H$ can be chosen as a random matrix,
        drawn from one of the Gaussian Ensembles GOE $(\beta=1)$, GUE
        $(\beta=2)$ or GSE $(\beta = 4)$, where $\beta= 1,4$ describe
        time--reversal invariant systems. Here, we focus on the
        time--reversal non--invariant unitary $(\beta=2)$ case of
        Hermitean $N\times N$ matrices $H$ drawn from a Gaussian
        probability density with variance $\nu^2/N$. The technically
        much more complicated cases $\beta=1,4$ will be treated in a
        forthcoming publication elsewhere.

	\textit{Ericson regime and Weisskopf estimate} ---
The Ericson
transition is governed by the ratio of average resonance
width $\Gamma$ and average mean level spacing $D$, \textit{i.e.} by the dimensionless parameter  
\begin{align}\label{Weiss}
	\Xi=\frac{1}{2\pi} \sum\limits_{c=1}^{M}T_c = \frac{\Gamma}{D}\ , 
\end{align}
where the second equation is known as the Weisskopf estimate \cite{Weisskopf} and $T_c$ are the transmission
coefficients. Importantly, the number $M$ of channels must be large, see $\textit{Appendix A}$ for details. 
        
	We have two objectives: First, we prove
        analytically that the real and imaginary parts of $S_{ab}$ for
        $a \neq b$ are Gaussian distributed in the Ericson regime. Up
        to now the reasoning was mainly heuristic
        \cite{AGASSI1975}. Second, we provide asymptotic corrections
        to the Gaussian form, thereby revealing the complete
        transition to the Ericson regime.
	
	\textit{Supersymmetry and starting point} --- 
	In Refs.~\cite{Efetov,VWZ,Efetov1996} a variant of the supersymmetry method was put forward
        \cite{PhysRevLett.111.030403, Matrixelemente,
          Wirkungsquerschnitte} that made it possible to explicitly
	calculate the distributions $P_s(x_s)$ of the real part, $x_1=\text{Re}(S_{ab})$, and the imaginary part, $x_2=\text{Im}(S_{ab})$, of $S_{ab}$
        and the cross section $\sigma_{ab}$ in terms of
        low--dimensional integrals. We employ those
        results to study the Ericson transition by means of asymptotic
        expansions in powers of $1/ \Xi$. Previously, only the
	correlator of the $S$ matrix for $\beta=1$ was
        investigated in this way \cite{VerbaarschotCorr}.

        \indent\textit{Derivation of the Universal Gaussian} --- The
        characteristic function generates the moments $\overline{x_s^n}$
        of the distribution $P_s(x_s)$, \textit{i.e.},
        \begin{align}
           R_s(k) = \int\limits_{-\infty}^{+\infty}dx_s P_s(x_s)e^{-ikx_s}
                            = \sum_{n=0}^\infty (-i)^n\frac{k^n}{n!}\overline{x_s^n} \ .
          \label{eq:sum} 
	\end{align}
        The exact analytical result for the characteristic function is
        given by~\cite{PhysRevLett.111.030403,Matrixelemente}
	\begin{align}\nonumber
	  R_s(k) =\ &1 -\int\limits_{1}^{\infty} d\lambda_1\int\limits_{-1}^{1} d\lambda_2
           \frac{k^2}{4 (\lambda_1 -\lambda_2)^2}\mathcal{F}_U(\lambda_1,\lambda_2)\\\label{eq:Sub1}
		&\times \left( t_a^1t_b^1+t_a^2t_b^2\right) J_0\left( k \sqrt{ t_a^1t_b^1}\right),\ s=1,2
	\end{align}
	where $t_c^j = \sqrt{|\lambda_j^2-1|}/(g_c^++\lambda_j), g_c^+
        = 2/T_c-1$ and $J_0$ denotes the Bessel function of
        order zero.  The channel factor reads
	\begin{align}
		\mathcal{F}_U(\lambda_1,\lambda_2) = \prod\limits_{c=1}^{M} \frac{g_c^+ + \lambda_2}{g_c^++\lambda_1} \ .
	\end{align}
	The characteristic functions for $s=1,2$ and thus their
        distributions are equal for $\beta=2$
        \cite{PhysRevLett.111.030403, Matrixelemente}.

        We expand the moments asymptotically in powers of $1/\Xi$. To
        this end, we expand $R_s(k)$ in Eq.~\eqref{eq:Sub1}, \textit{i.e.},
        $J_0\left( k \sqrt{ t_a^1t_b^1}\right)$ in powers of $k$. As
        $J_0$ is even in $k$, all odd moments vanish. All even moments
	necessarily exist due to the unitarity of the $S$
        matrix. The zeroth order moment equals one and is the
         Efetov--Wegner term \cite{Efetov, Efetov1996, Kieburg2009} in
	the supersymmetry calculation. The integrand in \cref{eq:Sub1} is 
	singular at $\lambda_1=\lambda_2=1$. This is typical for
        supersymmetry calculations, see Refs.~\cite{VWZ,
          Efetov,VerbaarschotCorr, PLUHAR19951,Efetov1996,Fyodorov,
          Rozhkov2003,Rozhkov2004,PhysRevLett.111.030403,
	  Matrixelemente, Wirkungsquerschnitte}, and entails that the computation of the remaining low-dimensional integrals is, in general, very challenging or even unfeasible. Yet, we succeeded in removing the singularity, with an exact change of variables $\lambda_1 = 1+q'r',\ \lambda_2 = 1-q'(1-r').$ As the
        integrand combined with the Jacobian contributes a factor of
        $q'^2$ this cancels the $(\lambda_1-\lambda_2)^{-2}=q'^{-2}$
	contribution at $q'=0$ and the integrand is finite. 
  We split the integration domain into two parts, $R_s(k)= R_s^{(d)}(k)+R_s^{(c)}(k)$, a disconnected
part, $R_s^{(d)}(k)$, with independent integration variables, and a connected
	part, $R_s^{(c)}(k)$, where
	\begin{align}
		R_s^{(d)}(k) &= 1 -\frac{k^2}{4}\int\limits_{0}^{2} dq'\int\limits_{0}^{1} dr' f(q',r';k)\label{Rsd}\\
		R_s^{(c)}(k) &=-\frac{k^2}{4}\int\limits_{2}^{\infty} dq'\int\limits_{(q'-2)/q'}^{1}  dr' f(q',r';k).\label{Rsc}
	\end{align}
The integrand is given by
	\begin{align}
	  & f(q',r'{};k) =  \mathcal{F}_U(1+q'r',1-q'(1-r'))\\\nonumber
          & \quad \times
	       \Biggl[ \frac{r'(q'r'+2)}{(g_a^++ q'r'+1)(g_b^+ +q'r'+1)}\\\nonumber
          & \quad +\frac{ (1-r')(2-q'(1-r'))}{(g_a^++1-q'(1-r'))(g_b^++1-q'(1-r'))}
                    \Biggr]\\\label{eq:Rk}&\times  J_0\left( k \sqrt{\frac{q' r'(q'r'+2)}{(g_a^++q'r'+1)(g_b^+ +q'r'+1)}}\right) \ .
	\end{align} 
	In a related but different context, the channel factor was
        approximated in Ref.~\cite{VerbaarschotCorr} for infinitely many
        channels with $T_c\simeq 1/M$ in the form
	\begin{align}
		\mathcal{F}_U(1+q'r',1-q'(1-r'))&\simeq \exp{(-\pi \Xi q' )} \ .
	\end{align} 
	We use this limit to calculate the moments. A more
        general channel factor expansion is given in $\textit{Appendix
          B}$. We obtain the asymptotes for the moments by
        employing Watson's lemma \cite{Expansions, Laplace1, Laplace2,
	  Wong} to express the integral over $q^\prime$ in terms of the derivatives of
        $f(q',r';k)/\mathcal{F}_U(1+q'r',1-q'(1-r'))$ with respect to
	$q^\prime$ at $q^\prime=0$ where only Eq.~(\ref{Rsd}) contributes. There, the
        exponentially decaying channel factor becomes
        maximal. We find
	\begin{align}\nonumber
	  \overline{x_s^{2n}} &= \frac{\Gamma{(2n+1)}}{2^{2n}\Gamma^2(n)}\\\nonumber
                             & \qquad \times \sum\limits_{m=0}^{\infty} \left(\frac{1}{\pi
            \Xi}\right)^{m+1} \frac{\partial^m}{\partial q'^m}
           \int\limits_{0}^{1}dr'(q' r')^{n-1}\\\notag
           &\times \Biggl[
            \frac{r'(q'r'+2)}{(g_a^++ q'r'+1)(g_b^+ +q'r'+1)}
            \\\nonumber&+\frac{
              (1-r')(2-q'(1-r'))}{(g_a^++1-q'(1-r'))(g_b^++1-q'(1-r'))}\Biggr]\\\label{eq:KorrekturMomente}&\times\left(
          \frac{q'r'+2}{(g_a^++q'r'+1)(g_b^+ +q'r'+1)} \right)^{n-1}
          \Biggl|_{q'=0} \ .
	\end{align}
	After swapping the order of differentiation and integration,
        we carry out the derivatives over $q'$. The remaining
	integrand is a polynomial in $r'$. Evaluating 
	their leading order term yields
	\begin{align}
		\overline{x_s^{2n}} 
		=\ & \frac{(2n)!}{n!} \left(\frac{1}{2(g_a^+ + 1)(g_b^+ + 1)\pi \Xi }\right)^n \ 
	\end{align}
	to order $\Xi^{-n}$.
	The explicit calculation is given in $\textit{Appendix C}$.
        Inserting the leading order asymptote into \cref{eq:sum} leads
        to $R_s(k)\simeq R_s^{(l)}(k)$, with
	\begin{align}\label{eq:erste}
		R^{(l)}_s(k) = \exp{\left(-\frac{k^2}{2(g_a^+ + 1)(g_b^+ + 1)\pi \Xi}\right)} \ .
	\end{align}
	By taking the inverse Fourier transform of the characteristic
        function, we arrive at the Gaussian approximation of the
        distribution $P_s(x_s)$ for either the real part $x_1$ or
        imaginary part $x_2$ of the off-diagonal $S$--matrix
        elements. Even though our derivation is not related to the
        Central Limit Theorem, the asymptotics necessarily requires a
        similar rescaling of $x_s$, namely $\xi_s = \sqrt{\Xi}x_s$.
        We find
	\begin{align}\nonumber
	  P_s^{(l)}(\xi_s) & \simeq \sqrt{\frac{ (g_a^+ + 1)(g_b^+ + 1)}{2}}\\
          \label{eq:Gaussian}&\times\exp{\left(-\frac{\pi (g_a^+ + 1)(g_b^++1)\xi_s^2}{2}\right)} 
	\end{align}
	for $s=1, 2$.  This proves the universal Gaussian,
        heuristically established by Ericson. The contribution from
        the connected part is globally decaying with $ \Xi$ and thus
        negligible. 
	
	\textit{Transition and higher order corrections} --- To
        subleading orders in $1/\Xi$, the moments are
	\begin{align}\nonumber
		\overline{x_s^{2n}}
		\simeq\ &\frac{(2n)!}{n!} \left(\frac{1}{2(g_a^+ + 1)(g_b^+ + 1)\pi \Xi }\right)^n\\
                +&\frac{(2n)!}{\Gamma(n-1)}  \frac{g_a^+g_b^+-g_a^+-g_b^+-3}{(2(g_a^++1)(g_b^++1)\pi \Xi)^{n+1}}
	\end{align}
		to order $\Xi^{-(n+1)}$.
        By resummation, the corresponding asymptote of the
        characteristic function becomes $R_s(k)\simeq
        R^{(l)}_s(k)+R^{(sl)}_s(k)$ with
	\begin{align}\nonumber
	  R_s^{(sl)}(k)\simeq\ & \exp{\left(-\frac{k^2}{2(g_a^++1)(g_b^++1)\pi \Xi}\right)}\\\label{eq:FourierChar}
          &\times\frac{(g_a^+g_b^+-g_a^+-g_b^+-3)k^4}{8((g_a^++1)(g_b^++1)\pi \Xi)^3} \ .
	\end{align}
	The associated correction $P_s^{(sl)}(\xi_s)$ to the rescaled
        distribution, $P_s(\xi_s)\simeq
        P_s^{(l)}(\xi_s)+P_s^{(sl)}(\xi_s)$ is thus
	\begin{align}\label{eq:Zweite} \nonumber
        P_s^{(sl)}(\xi_s)\simeq& \sqrt{\frac{ (g_a^+ + 1)(g_b^+ +
            1)}{2}}\\\nonumber
            &\times \exp{\left(-\frac{\pi (g_a^+ +
            1)(g_b^++1)\xi_s^2}{2}\right)}\\\nonumber&\times\Biggl(3-6\pi
          (g_a^+ + 1)(g_b^+ +1)\xi_s^2\\\nonumber&+(\pi (g_a^+ +
          1)(g_b^+ + 1))^2\xi_s^4\Biggr) \\&\times
          \frac{g_a^+g_b^+-g_a^+-g_b^+-3}{8(g_a^++1)(g_b^++1)\pi \Xi}
	\end{align}
	which goes with $1/\Xi$. For $\Xi\gg1$, the distribution
        $P_s(\xi_s)\simeq P_s^{(l)}(\xi_s)+P_s^{(sl)}(\xi_s)$
        converges to a Gaussian. For details of the derivation, see
        $\textit{Appendix D}$. The asymptotic approximation as well as
        the symmetry $a\leftrightarrow b$ of the scattering channels
        comply with the symmetry $P_s(\xi_s)=P_s(-\xi_s)$. The
        correction to the Gaussian maximum at $\xi_s=0$ is
	\begin{align}
	  P_s^{(sl)}(0)= &\frac{3(g_a^+g_b^+-g_a^+-g_b^+-3)}{8\pi\Xi\sqrt{2(g_a^++1)(g_b^++1)}} 
          \label{eq:CorrMax}
	\end{align}
	from \cref{eq:Zweite}.
	Thus, for $\Xi \approx 1$, \textit{i.e.}, for weakly--overlapping
        resonances, the distribution is not Gaussian, rather it is
        shifted by the quantity $P_s^{(sl)}(0)$ that depends on the
        transmission coefficients $T_a$ and $T_b$. This shift is
        positive if $1-T_a-T_b>0$ and negative otherwise, cf.~\cref{eq:CorrMax}.
	
	\textit{Cross--section distribution} --- In many experiments
        only the cross sections are accessible. We may calculate the
        distribution of
        $\sigma_{ab}=\text{Re}^2S_{ab}+\text{Im}^2S_{ab}$ via a
        two--dimensional Fourier transform from a bivariate
        characteristic function \cite{Wirkungsquerschnitte}. Due to
        the radial symmetry of the latter, we can write the
        cross--section distribution as Hankel transform of the
        univariate characteristic function,
	\begin{align}
	  p(\sigma_{ab}) = \frac{1}{2}\int\limits_0^\infty {\rm d}k\ kR_s(k)J_0(\sqrt{\sigma_{ab}}k) \ .
        \label{Hankel}
	\end{align}
	Using $R_s(k)\simeq R^{(l)}_s(k)+R^{(sl)}_s(k)$ in
        Eqs. (\ref{eq:erste}) and~(\ref{eq:FourierChar}) and rescaling
        $\widetilde{\sigma}_{ab}=\sigma_{ab}/\langle
        \sigma_{ab}\rangle$, we arrive at
        $p(\widetilde{\sigma}_{ab})\simeq
        p^{(l)}(\widetilde{\sigma}_{ab})+p^{(sl)}(\widetilde{\sigma}_{ab})$,
        with
	\begin{align}
	p^{(l)}(\widetilde{\sigma}_{ab}) &= \exp{(-\widetilde{\sigma}_{ab})} \notag\\
        p^{(sl)}(\widetilde{\sigma}_{ab})&= \frac{\exp{(-\widetilde{\sigma}_{ab})}((4-\widetilde{\sigma}_{ab})\widetilde{\sigma}_{ab}-2) }{2\pi\Xi(g_a^++1)(g_b^++1)}\notag\\
        &\times(3+g_a^++g_b^+-g_a^+g_b^+) \ .
        \label{crosssectionasymp} 
	\end{align}
	The rescaling amounts to normalizing the cross--section
        in the distribution to its expectation value in the Ericson regime,
        given by $\langle \sigma_{ab} \rangle =
        2/(\pi\Xi(g_a^++1)(g_b^++1))$ to order $\Xi^{-1}$. There is
        an additional contribution to the pure exponential decay in
        the Ericson regime. At $\widetilde{\sigma}_{ab}=0$, we have
	\begin{align}
	   p(0)=1-\frac{3+g_a^++g_b^+-g_a^+g_b^+}{\pi\Xi(g_a^++1)(g_b^++1)}  \ .
        \label{pvon0}
        \end{align}
	The second term gives the shift
	relative to the maximum of the exponential decay. This quantitatively explains the
        observed deviations from the exponential decay in the onset of
	the Ericson regime \cite{Wirkungsquerschnitte}. The distribution of cross sections is closely related to that of the wave-function intensities $I$, $\mathcal{P}_M(I)$, in an open wave-chaotic cavity with violated time--reversal invariance~\cite{FyodorovSafonova}. By means of a similar analysis as presented above, asymptotes for $\mathcal{P}_M(I)$ are derived in the Supplemental Material \cite{supp}. The leading--order term yields an exponential decay of $\mathcal{P}_M(I)$ in the limit $\Xi\gg1$.
	
	\indent\textit{Validation with experimental data and numerical
          simulations} --- We compare our results to data from
        measurements with microwave networks
        \cite{Quantengraphen}. These were designed such that they
	simulate open quantum graphs~\cite{QuantumChaos,QuantumChaos2} with violated time-reversal invariance~\cite{Hul2005}
	exhibiting quantum chaotic scattering~\cite{ZM13} for $\beta=2$.  The
        Heidelberg approach applies and numerical studies revealed that the typical 
	values of the transmission coefficients $T_c$ and of $M$ in such microwave experiments 
	suffice to ensure applicability of the Weisskopf estimate~\cite{DietzWeidenmueller}.
	We have two strong scattering
        channels $a=2, b=1$ with $T_1=T_2=\num{0.967}$ and a parameter
        $\tau_{\text{abs}}=7.013$ to account for absorption
        \cite{Dietz2010}. To model it, $50$ fictitious channels of
        equal transmission were used. Using the Weisskopf estimate $
        \Xi=(T_1+T_2+\tau_{\text{abs}})/(2\pi) \approx \num{1.424} $
        we find that the measurements are taken in the onset of the
        Ericson regime, the resonances are weakly overlapping. We
        focus on the $S$--matrix element $S_{21}$.
	\begin{figure}[htbp]
		\centering
		\captionsetup[subfigure]{labelformat=empty}
		\begin{subfigure}[b]{1\linewidth}
			\centering
			\includegraphics[width=\linewidth]{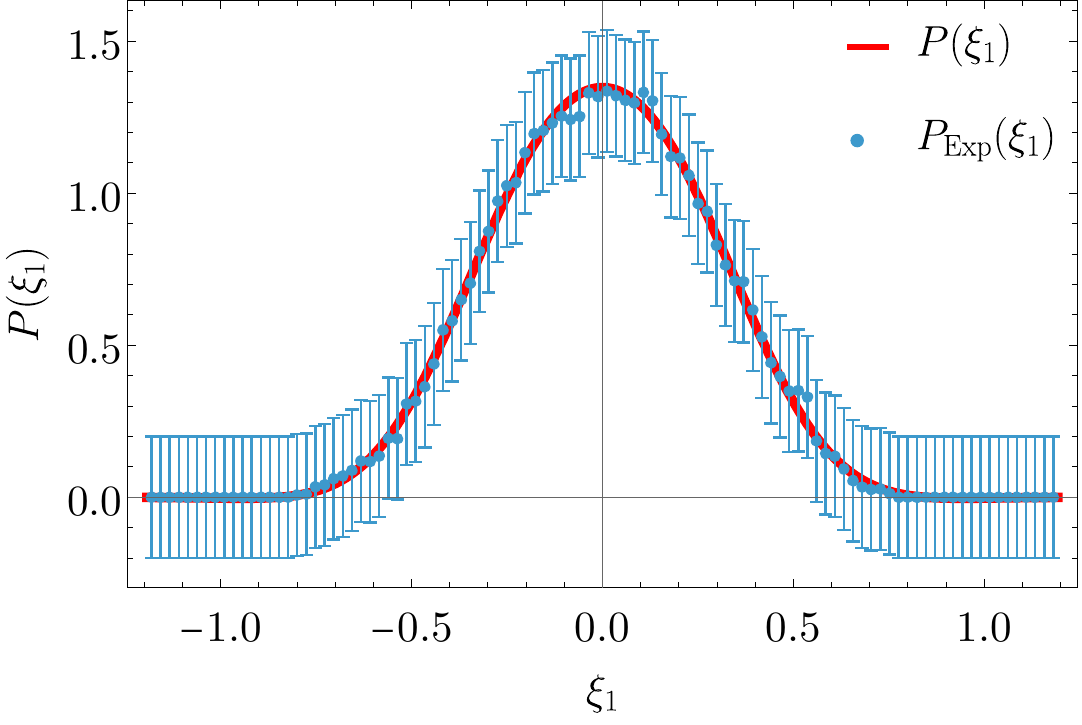} 
			\subcaption{}
		\end{subfigure}
		~
		~	
		\begin{subfigure}[b]{1\linewidth}
			\centering
			\includegraphics[width=1\linewidth]{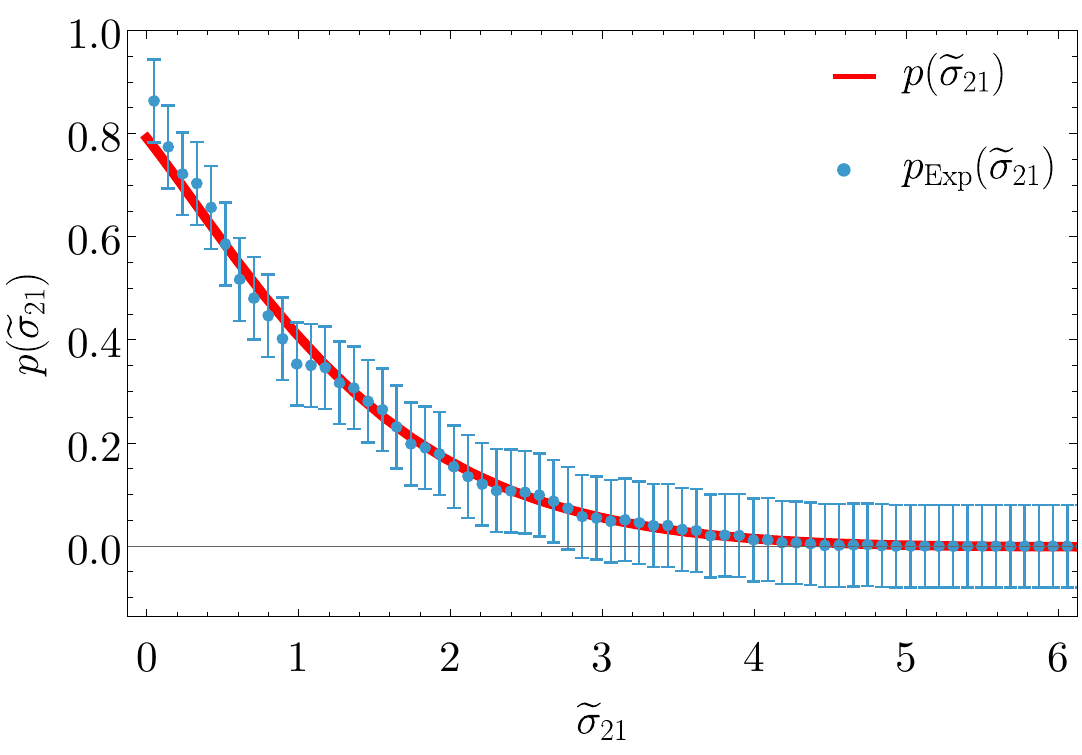} 
			\subcaption{}
		\end{subfigure}	
		\caption{Experimental and analytical results for the
		  distributions of rescaled 
		  $S$--matrix elements
                  (top) and of cross sections (bottom) for $\Xi=1.424$
                  and $M=52$. Blue error bars indicate the empirical
                  standard deviation around each point of
                  measurement.}
		\label[figure]{Distributions1}
	\end{figure}
	As seen in \cref{Distributions1}, the analytical result is in
        very good agreement with the experimental data. The measured
        distribution is equal for real and imaginary part. The peak of
        the distribution is non--Gaussian, as to be expected in the
        onset of the Ericson regime.

\begin{figure}[htbp]
		\centering
		\captionsetup[subfigure]{labelformat=empty}
		\begin{subfigure}[b]{1\linewidth}
			\centering
			\includegraphics[width=\linewidth]{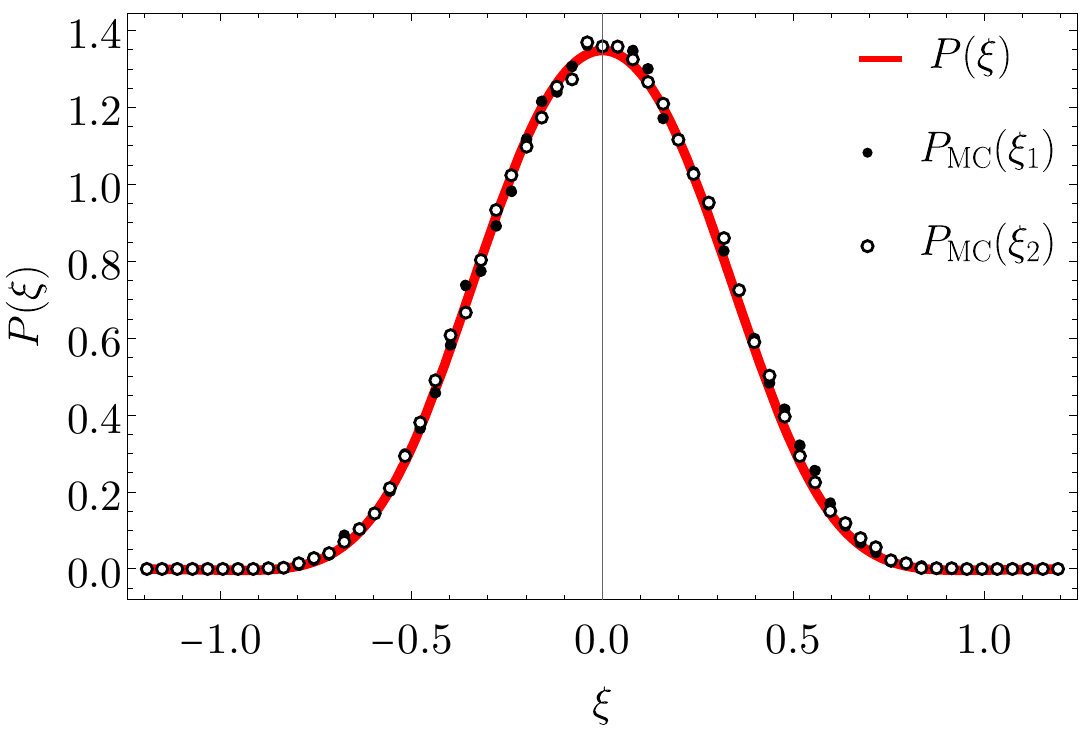} 
			\subcaption{}
		\end{subfigure}
		~
		\begin{subfigure}[b]{1\linewidth}
			\centering
			\includegraphics[width=\linewidth]{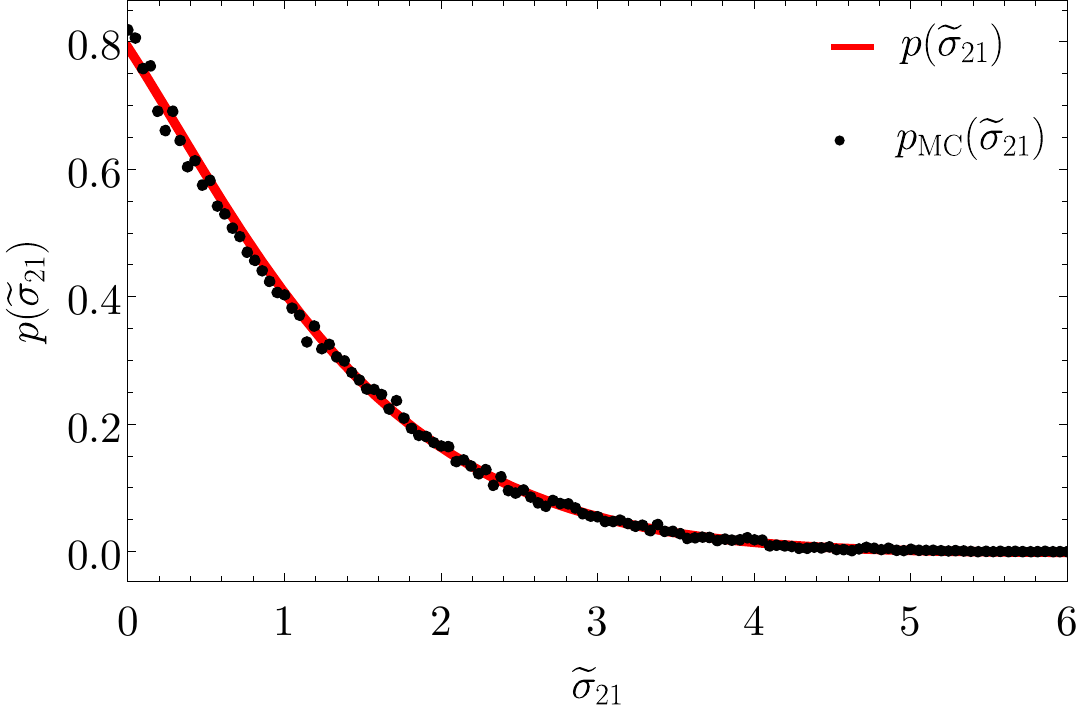} 
			\subcaption{}
		\end{subfigure}
		\caption{Monte Carlo simulations and analytical
		  results for the distributions of rescaled 
		  $S$--matrix elements 
		  (top) and of cross sections (bottom)
                  for $\Xi=1.424$ and $M=52$.}
		\label[figure]{DistributionsMC}
	\end{figure}

\begin{figure}[htbp]
		\centering
		\captionsetup[subfigure]{labelformat=empty}
		\begin{subfigure}[b]{1\linewidth}
			\centering
			\includegraphics[width=\linewidth]{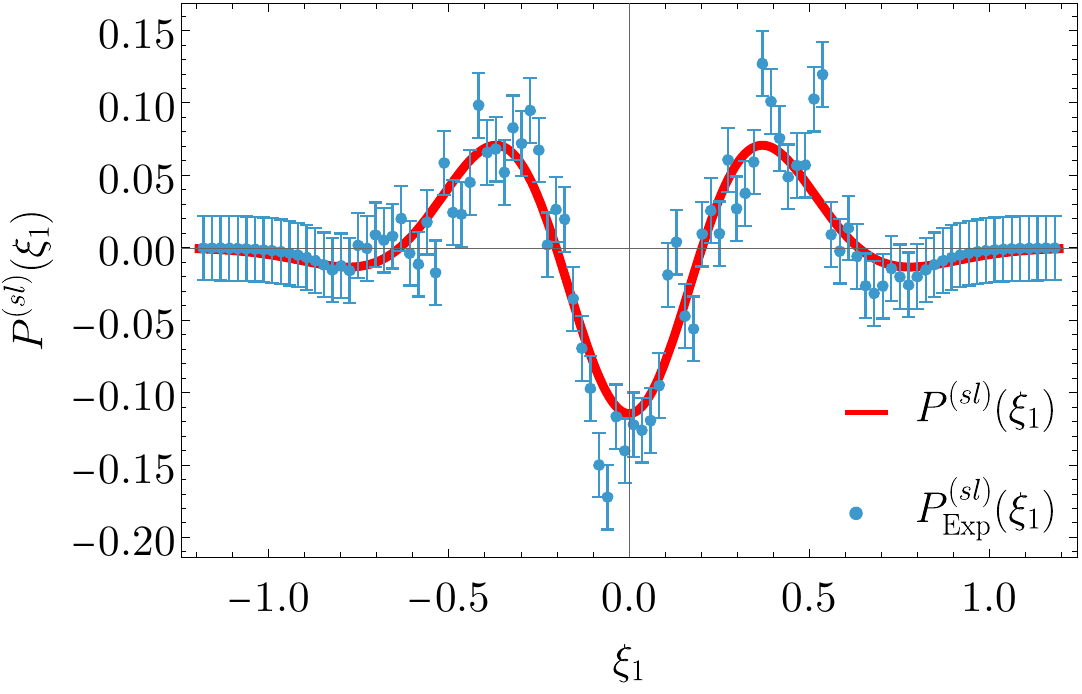} 
			\subcaption{}
		\end{subfigure}
		~
		\begin{subfigure}[b]{1\linewidth}
			\centering
			\includegraphics[width=1\linewidth]{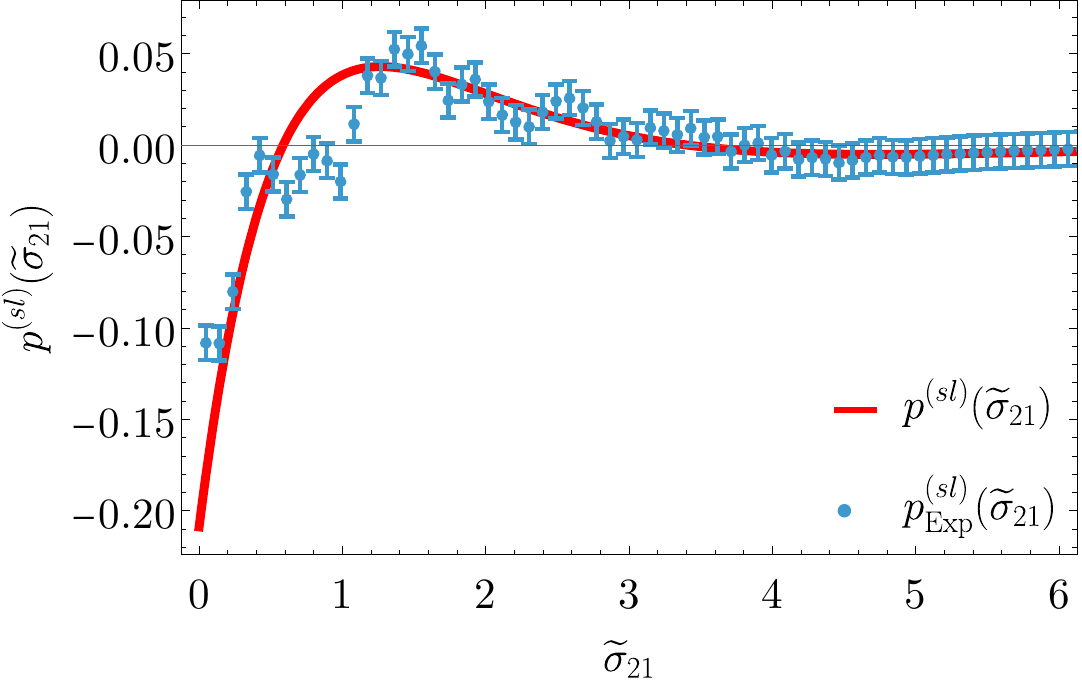} 
			\subcaption{}
		\end{subfigure}	
		\caption{Experimental data and the subleading terms
                  for the distributions of rescaled 
		  $S$--matrix
                  elements (top) and of cross sections (bottom) for
                  $\Xi=1.424$ and $M=52$.}
		\label[figure]{Corrections}
	\end{figure}

        We also compare the analytical result with Monte Carlo
        simulations which were done by sampling $30000$ random
        matrices with dimension $N=200$ from the GUE and numerically
	calculating the corresponding $S$--matrix
	elements. As
        displayed in \cref{DistributionsMC}, the agreement is very good.
	To visualize the transition to the Gaussian distribution
        $P_s^{(l)}(\xi_s)$ in the Ericson regime, we depict in
        \cref{Corrections} the subleading terms $P^{(sl)}_s(\xi_1)$
        and $p^{(sl)}(\widetilde{\sigma}_{ab})$ for the distributions
        in comparison with the experimental data.
	The sizeable dip at zero in \cref{Corrections} demonstrates
        the significant deviations from the Gaussian for weakly
        overlapping resonances. The distribution is also broadened in
	the tails.  The cross--section distribution 
 		\begin{figure}[htbp]
		\centering
		\captionsetup[subfigure]{labelformat=empty}
		\begin{subfigure}[b]{1\linewidth}
			\centering
			\includegraphics[width=\linewidth]{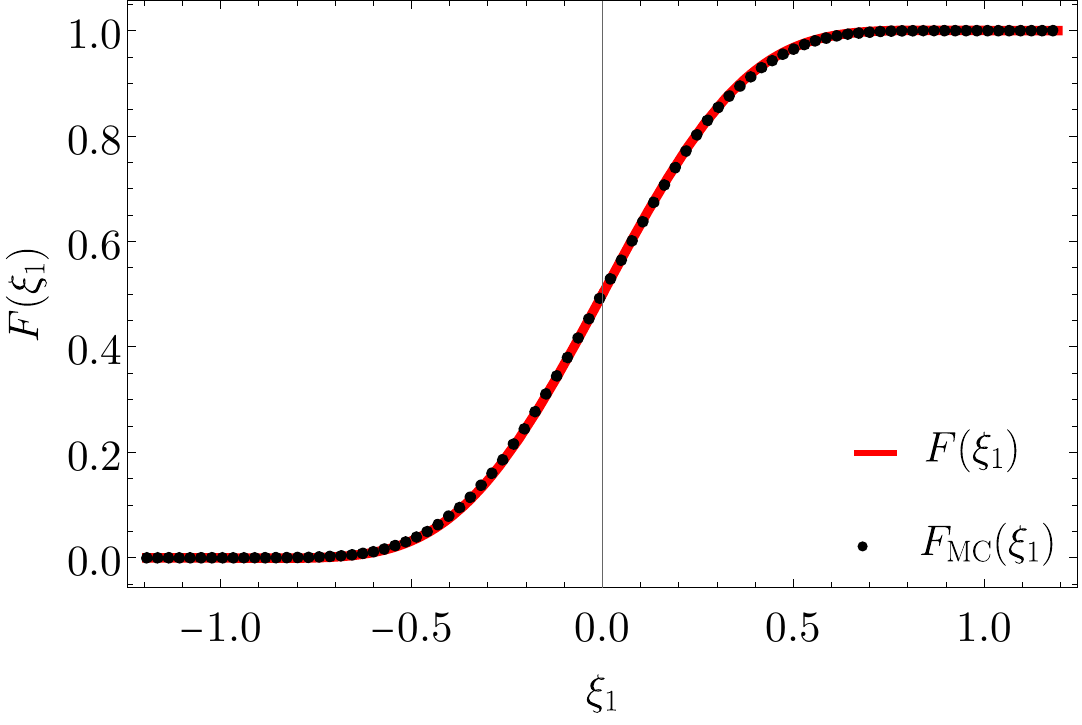} 
			\subcaption{}
		\end{subfigure}
		~
		\begin{subfigure}[b]{1\linewidth}
			\centering
			\includegraphics[width=1\linewidth]{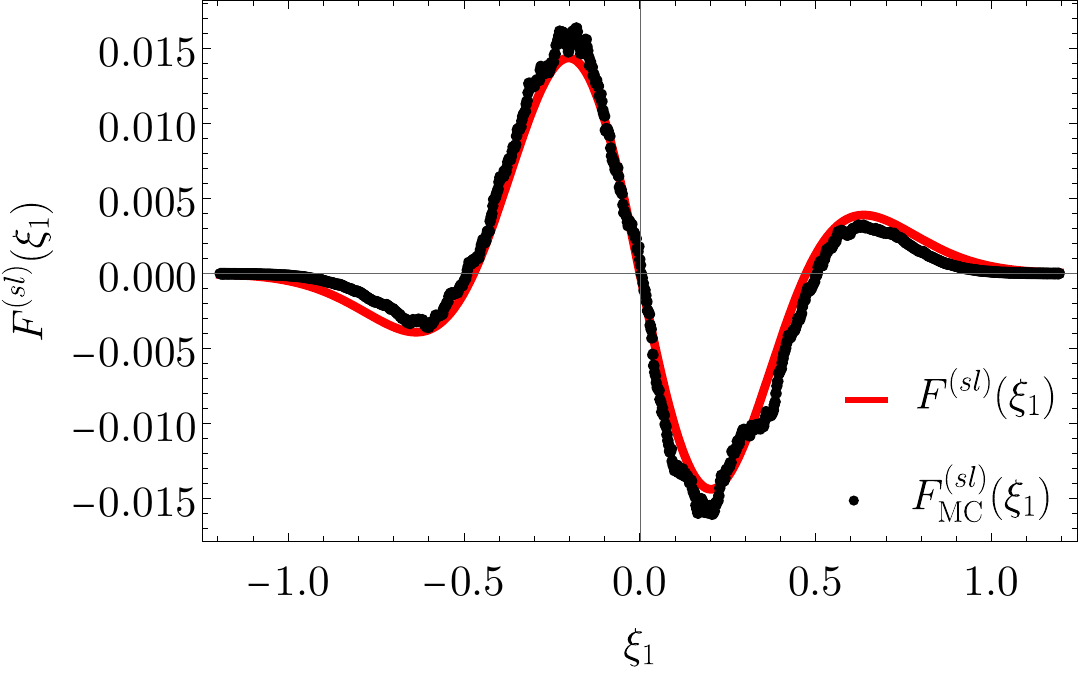} 
			\subcaption{}
		\end{subfigure}	
		\caption{Full cumulative distribution $F(\xi_1)$ of
		the rescaled $S$--matrix elements (top) and subleading
                  term $F^{(sl)}(\xi_1)$ (bottom) for $\Xi=1.424$ and $M=52$.}
		\label[figure]{CDFs}
	        \end{figure}
        shows a maximal deviation from the exponential decay at zero,
        which is clearly visible in \cref{Corrections}. To numerically
        test the subleading corrections, we use the cumulative
        distribution function (CDF) $F(\xi_s)
        =F^{(l)}(\xi_s)+F^{(sl)}(\xi_s)$ in \cref{CDFs}. Since
        $P_s(\xi_s)$ is symmetric around $\xi_s=0$, $F(\xi_s)$ and
        $F^{(sl)}(\xi_s)$ are antisymmetric in $\xi_s$, see \cref{CDFs}.

        An important comment is in order. One might question that our
        asymptotic results are applicable for the relatively small
	value of $\Xi=1.424$. Numerical simulations
        of the full problem compared to the analytically obtained
	subleading term are displayed in \cref{CDFs}. The deviation is
	\begin{figure}[htbp]
		\centering
		\captionsetup[subfigure]{labelformat=empty}
		\begin{subfigure}[b]{1\linewidth}
			\includegraphics[width=1\linewidth]{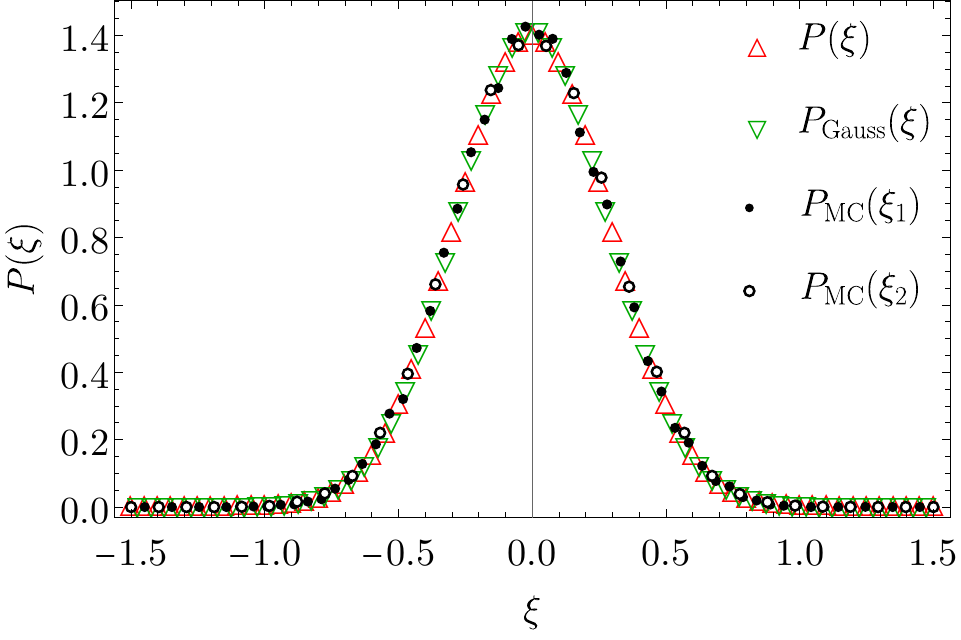} 
			\subcaption{}
		\end{subfigure}
		~
		\begin{subfigure}[H]{1\linewidth}
			\includegraphics[width=1\linewidth]{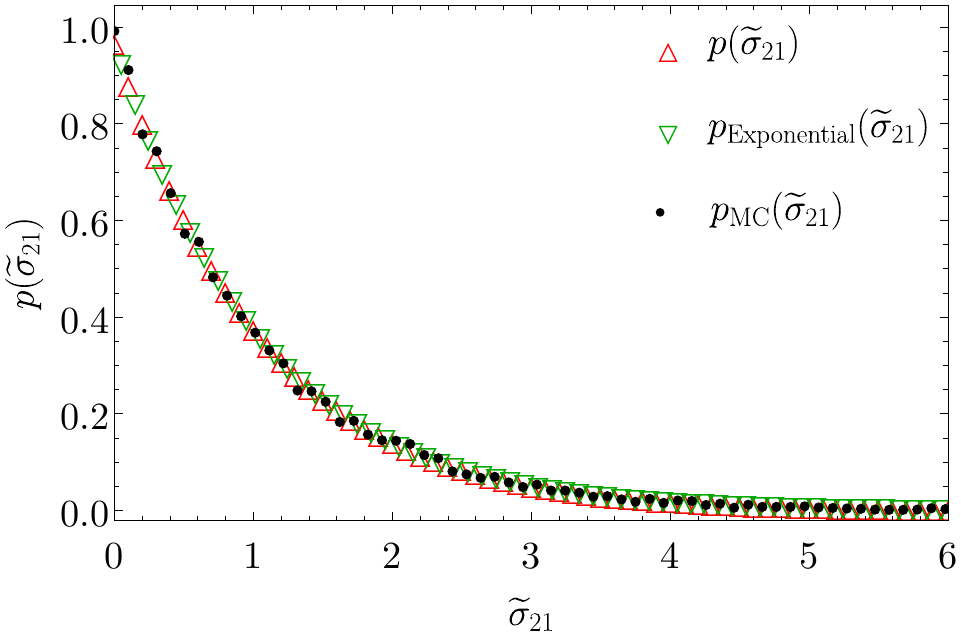} 
			\subcaption{}
		\end{subfigure}
		\caption{Analytical results and Monte Carlo
                  simulations for the distributions $P(\xi_1)$,
                  $P(\xi_2)$ of rescaled real and imaginary parts of
		  $S$--matrix elements (top) and for the
                  distribution $p(\widetilde{\sigma}_{21})$ of the
                  rescaled cross--sections (bottom) for $\Xi=9.55$ and
                  $M=60$.}
		\label[figure]{Gauss}
	\end{figure}
        very small, strongly indicating that, even for this small
        value of $\Xi$, the higher order terms in $1/\Xi$ are of little
        relevance. Put differently, the transition to the Ericson regime is
	so fast that in typical situations the subleading term  provides the relevant corrections.  To further illustrate this
        important point, we numerically test our results for a set of
        parameters deep in the Ericson regime. We choose
        $T_c=1$ for $M=60$ channels, which corresponds to $\Xi\approx
        9.55$. By doing so, the corrections in $1/\Xi$ become small
        compared to the leading order term and we indeed find Gaussian
        and exponential behavior, respectively, as depicted in
        \cref{Gauss}.
      
	\textit{Conclusion} --- We succeeded in fully solving a
        longstanding problem in quantum chaotic or stochastic
        scattering. We quantitatively captured the Ericson transition
        of time--reversal non--invariant systems ($\beta=2$) within
        the Heidelberg approach, and derived the universal Gaussian
        distribution for the Ericson regime that emerges in the
        universal stochasticity of the scattering
        problem. The Heidelberg approach as such already incorporates universality as the random matrices modeling the Hamiltonian $H$ in the interaction zone represent in a broad sense a generically chaotic or complex system. On top of this universality emerges another universality, \textit{i.e.} the Gaussians, in the Ericson transition. Furthermore, we provided a full information on the
        transition to this Ericson limit by an asymptotic analysis. To
        this end, we calculated all moments for the probability
        distribution of the off--diagonal $S$--matrix elements and
        cross sections.  Removing a cumbersome singularity was crucial.
        We carried out detailed validations with experimental data and
        numerical Monte Carlo simulations. The agreement with our
        analytical results is very good. We also found that the
        Ericson transition is very fast in the sense that the
        subleading term in the asymptotic expansion will typically
        suffice to understand deviations from the universal Gaussian
        in experimental data.  The methods developed here carry over
        to the case of orthogonally and symplectically invariant
        Hamiltonians, \textit{i.e.}, to the time--reversal invariant cases
        $\beta=1$ and $\beta=4$.  As the calculations become
        technically much more involved, this material is beyond the
        scope of this contribution and will be presented
        elsewhere.

        \textit{Acknowledgments} --- We thank Nils Gluth for fruitful
        discussions. Two of us (TG, SK) are grateful to the Deutsche
        Forschungsgemeinschaft (DFG, German Research Foundation) for
        support within the project Stochastic Quantum Scattering --
        New Tools, New Aspects, DFG project number 540160740.  Two of
        us (JC, BD) acknowledge financial support for the experiments
        from the China National Science Foundation (NSF), grants
        11775100, 12247101 and 11961131009. One of us (BD) is grateful
        for funding from the Institute for Basic Science in Korea,
        project IBS–R024–D1.

\twocolumngrid
\bibliography{EricsonTransition}
\appendix
\onecolumngrid
\section{Endmatter}
\textit{Appendix A: Moldauer--Simonius relation and Weisskopf estimate} --- 
	The Ericson regime is characterized by two prominent features. Resonances are strongly overlapping, that is, the ratio $\Gamma/D$ of the average resonance width $\Gamma$ and average mean level spacing $D$ is large. Furthermore, the number $M$ of open channels with associated transmission coefficients $T_c$ is large. A relation between the transmission coefficients and $\Gamma/D$ is given by the Moldauer-Simonius relation~\cite{Simonius,PhysRevC.11.426,FyodorovSommersGamma,FyodorovJMP1997,KottosSchanz,DietzWeidenmueller}
\begin{align}
	\frac{\Gamma}{D}= -\frac{1}{2\pi}\sum\limits_{c=1}^{M}\ln(1-T_c)\label{eq:Simonius}
\end{align}
which was derived rigorously for $\beta=2$ in \cite{FyodorovJMP1997}.
For $T_c \ll1,\ M\gg1$, \cref{eq:Simonius} yields  
for the dimensionless parameter                
\begin{align}
	\Xi = \frac{1}{2\pi}\sum\limits_{c=1}^{M}T_c
\end{align}
the Weisskopf estimate 
\begin{align}
	\frac{\Gamma}{D} = \Xi.
\end{align}
It provides a good approximation also for a large, but finite number of open channels and for larger values of $T_c$~\cite{DietzWeidenmueller}.

\textit{Appendix B: Channel factor expansion} --- We write the channel factor as exponential and identify the parametric dependencies in $\Xi$ and $T_c$,
	\begin{align}
	F_U(\lambda_1,\lambda_2)&\nonumber= \prod\limits_{c=1}^{M }\frac{g_c^++\lambda_2}{g_c^++\lambda_1}=\exp{\ln{\left(	\prod\limits_{c=1}^{M }\frac{g_c^++\lambda_2}{g_c^++\lambda_1}\right)}} = \exp{\left(\sum\limits_{c=1}^{M}\ln{\left(\frac{1+\frac{T_c}{2}(\lambda_2-1)}{1+\frac{T_c}{2}(\lambda_1-1)}\right)}\right)}\\&=\exp{\left(\sum\limits_{c=1}^{M}\sum\limits_{k=1}^\infty\frac{(-1)^{k+1}}{k}\Biggl(\Biggl(\frac{T_c}{2}(\lambda_2-1)\Biggr)^k-\Biggl(\frac{T_c}{2}(\lambda_1-1)\Biggr)^k\Biggr) \right)}.
\end{align}
In the last equation, we employed the Taylor series of the logarithm. Upon substituting	$\lambda_1= 1+q'r',\ \lambda_2= 1-q'(1-r')$ with $q'\in[0,2]$ and $r' \in [0,1]$, we arrive at
\begin{align}
	F_U(1+q'r',1-q'(1-r'))&\nonumber=	\exp{\left(\sum\limits_{c=1}^{M}\sum\limits_{k=1}^\infty\Biggl(\frac{T_c}{2}\Biggr)^k\frac{(-1)^{k+1}}{k}\Biggl((q'(r'-1))^k - (q'r')^k\Biggr) \right)}\\\nonumber&=
	\exp{\left(-\sum\limits_{c=1}^{M}\frac{T_c}{2}q' \right)}\exp{\left(\sum\limits_{c=1}^{M}\sum\limits_{k=2}^\infty\Biggl(\frac{T_c}{2}\Biggr)^k\frac{(-1)^{k+1}}{k}\Biggl((q'(r'-1))^k - (q'r')^k\Biggr) \right)}\\&=\exp{\left(-\pi \Xi q'\right)}\exp{\left(\sum\limits_{c=1}^{M}\sum\limits_{k=2}^\infty\Biggl(\frac{T_c}{2}\Biggr)^k\frac{(-1)^{k+1}}{k}\Biggl((q'(r'-1))^k - (q'r')^k\Biggr) \right)} \ .
\end{align}
From the exact expression above we can infer various regimes. The strongest case of the Ericson regime is given for $T_c=1$ and large $M$, resulting in the Gaussian and exponential form of the aforementioned distributions.
Moreover, we consider the case of many channels with small transmission coefficients. Now, when $T_c\simeq1/M$, we get in the large $M$-limit
\begin{align}
	\lim_{M \to \infty}&\nonumber\exp{\left(-\pi \Xi q'\right)}\exp{\left(\sum\limits_{c=1}^{M}\sum\limits_{k=2}^\infty\Biggl(\frac{T_c}{2}\Biggr)^k\frac{(-1)^{k+1}}{k}\Biggl((q'(r'-1))^k - (q'r')^k\Biggr) \right)}\\\nonumber&\simeq 	\lim_{M \to \infty}	\exp{\left(-\pi \Xi q'\right)}\exp{\left(\sum\limits_{c=1}^{M}\sum\limits_{k=2}^\infty\Biggl(\frac{1}{2M}\Biggr)^k\frac{(-1)^{k+1}}{k}\Biggl((q'(r'-1))^k - (q'r')^k\Biggr) \right)}\\&=\exp{\left(-\pi \Xi q'\right)} \ ,
\end{align}
since the sum over the transmission coefficients scales linearly with $M$. Thereby we recover the result of \cite{VerbaarschotCorr}. The transmission coefficients are not necessarily equal for all channels.
We conclude that only the term $\exp{\left(-\pi \Xi q'\right)}$ significantly contributes in the large $M$-limit of many small channels. However, if one has a few strong channels $T_c \simeq 1$ and $\Xi\simeq1$, terms including powers $T_c^j$, $j\geq 2$ contribute to the asymptotic expansion in higher order corrections due to the occurrence of $(q'T_c/2)^j$. Therefore, the leading and subleading order terms of the asymptotic expansion in $1/\Xi$ are generally essential for an arbitrary number of channels, thereby corroborating the relevance of the analytical expansion presented in this Letter. 

\textit{Appendix C: Explicit derivation of the  Gaussian} ---
Here, we explicitly calculate the integrals and derivatives arising in the leading order term 
of the 2n-th moment. Splitting the integral from \cref{eq:KorrekturMomente} into two parts and only calculating the leading order term ($m=n-1$), we have
\begin{align}\nonumber
	\int\limits_{0}^{1}dr'\frac{\partial^{n-1}}{\partial q'^{n-1}} q'^{n-1}\left(\frac{r'(q'r'+2)}{(g_a^+ + q'r'+1)(g_b^+ + q'r'+1)}\right)^n \Biggl|_{q'=0} &= \int\limits_{0}^{1}dr'2^n (n-1)!\left( \frac{r'}{(g_a^+ +1)(g_b^+ +1)}\right)^n \\&= 2^n (n-1)!\left( \frac{1}{(g_a^+ +1)(g_b^+ +1)}\right)^n \frac{1}{n+1} \ .
\end{align}
The second integral reads
\begin{align}\nonumber
	&\int\limits_{0}^{1}dr'\frac{\partial^{n-1}}{\partial q'^{n-1}} q'^{n-1}\Biggr[ \left( \frac{r'(q'r'+2)}{(g_a^+ + q'r'+1)(g_b^+ + q'r'+1)}        \right)^{n-1} \\\nonumber&\ \ \ \times\left(\frac{(1-r')(2-q'(1-r'))}{(g_a^+ +1-q'(1-r'))(g_b^++1-q'(1-r'))}\right)\Biggr] \Biggl|_{q'=0}\\\notag&= \int\limits_{0}^{1}dr' (n-1)! \left(\frac{2r'}{(g_a^+ + 1)(g_b^+ + 1)}\right)^{n-1}\frac{2(1-r')}{(g_a^+ + 1)(g_b^++1)}\\
	&= 2^n (n-1)! \left(\frac{1}{(g_a^+ + 1)(g_b^+ + 1)}\right)^n \left(   \frac{1}{n} -\frac{1}{n+1}     \right)=  2^n (n-1)! \left(\frac{1}{(g_a^+ + 1)(g_b^+ + 1)}\right)^n\frac{1}{n(n+1)} \ .
\end{align}
All in all, the leading order asymptotic expression for the moments is
\begin{align}\nonumber
	\overline{x_s^{2n}} =\ & \frac{\Gamma{(2n+1)}}{2^{2n}\Gamma^2(n)} \left(\frac{2}{2\pi \Xi}\right)^{n}\Biggl( 2^n (n-1)!\left( \frac{1}{(g_a^+ +1)(g_b^+ +1)}\right)^n \frac{1}{n+1}\\\notag& +2^n (n-1)! \left(\frac{1}{(g_a^+ + 1)(g_b^+ + 1)}\right)^n\frac{1}{n(n+1)}\Biggr)\\=\ &  \frac{\Gamma{(2n+1)}}{\Gamma(n)n}\left(\frac{1}{2(g_a^+ + 1)(g_b^+ + 1)\pi \Xi }\right)^n= \frac{(2n)!}{n!} \left(\frac{1}{2(g_a^+ + 1)(g_b^+ + 1)\pi \Xi }\right)^n \ .
\end{align}
By using the moment generating property of the characteristic function, we arrive at
\begin{align}\nonumber
	R_s^{(l)}(k) &= \sum\limits_{n=0}^{\infty}(-1)^n k^{2n} \frac{1}{(2n)!} \frac{(2n)!}{n!} \left(\frac{1}{2(g_a^+ + 1)(g_b^+ + 1)\pi \Xi }\right)^n\\\label{eq:ersted}
	&= \sum\limits_{n=0}^{\infty}(-1)^n k^{2n}\frac{1}{n!} \left(\frac{1}{2(g_a^+ + 1)(g_b^+ + 1)\pi \Xi }\right)^n = \exp{\left(-\frac{k^2}{2(g_a^+ + 1)(g_b^+ + 1)\pi \Xi}\right)} \ .
\end{align}
Performing the inverse Fourier transform yields the Gaussian asymptote
\begin{align}
\mathcal{F}^{-1}[R_s^{(l)}(k)](x_s) =\sqrt{\frac{ \Xi(g_a^++1)(g_b^++1)}{2}}\exp{\Biggl(-\frac{\pi \Xi(g_a^++1)(g_b^++1)x_s^2}{2}\Biggr)} \ .
\end{align}
 Higher order corrections are similarly calculated.

\textit{Appendix D: Unscaled distributions $P_s(x_s)$ and $p(\sigma_{ab})$} ---
The leading order terms of the unscaled distribution $P_s(x_s)\simeq P_s^{(l)}(x_s)+P_s^{(sl)}(x_s)$ are found to be 
\begin{align}\nonumber
P_s^{(l)}(x_s)&=\mathcal{F}^{-1}[R_s^{(l)}(k)](x_s) \\\nonumber&=\  \sqrt{\frac{ \Xi(g_a^++1)(g_b^++1)}{2}}\exp{\Biggl(-\frac{\pi \Xi(g_a^++1)(g_b^++1)x_s^2}{2}\Biggr)} \ ,\\\nonumber	P_s^{(sl)}(x_s)&\simeq\ \mathcal{F}^{-1}[R_s^{(sl)}(k)](x_s)\\
	\nonumber&=\ \sqrt{\frac{ \Xi(g_a^+ + 1)(g_b^+ + 1)}{2}}\exp{\left(-\frac{\pi \Xi(g_a^+ + 1)(g_b^++1)x_s^2}{2}\right)}\\
	\nonumber&\qquad \times\Biggl[3-6\pi \Xi(g_a^+ + 1)(g_b^+ +1)x_s^2+(\pi \Xi(g_a^+ + 1)(g_b^+ + 1))^2x_s^4\Biggr] \\&\qquad \times \frac{g_a^+g_b^+-g_a^+-g_b^+-3}{8(g_a^++1)(g_b^++1)\pi \Xi} \ .
\end{align}
The unscaled asymptotes consists of a Gaussian multiplied with a series in non-integer powers of $\Xi$, calling for an appropriate rescaling as done in the main text.
At $x_s=0$, we have
\begin{align}
	P_s^{(sl)}(0)=  \frac{3(g_a^+g_b^+-g_a^+-g_b^+-3)}{8\sqrt{2(g_a^++1)(g_b^++1)\pi \Xi}} \ .
\end{align}
We see that the first correction to the Gaussian maximum is of order $1/\sqrt{\Xi}$. The correction becomes negligible in the limit of large $\Xi$, which yields Gaussian behavior of the distribution. 
The leading order terms for the unscaled cross-section distribution $p(\sigma_{ab}) \simeq p^{(l)}(\sigma_{ab})+p^{(sl)}(\sigma_{ab})$ read
\begin{align}
			p^{(l)}(\sigma_{ab})&\nonumber=
		\frac{1}{2}\exp{\left(-\frac{\sigma_{ab} (g_a^++1)(g_b^++1)\pi \Xi}{2}\right)}(g_a^++1)(g_b^++1)\pi \Xi \ ,\\\nonumber
				p^{(sl)}(\sigma_{ab})&=\nonumber
			\frac{1}{2}\exp{\left(-\frac{\sigma_{ab} (g_a^++1)(g_b^++1)\pi \Xi}{2}\right)}\\\ &\qquad \times(g_a^+g_b^+-g_a^+-g_b^+-3)\Biggl(1-(g_a^++1)(g_b^++1)\pi \Xi \sigma_{ab} +\frac{1}{8}((g_a^++1)(g_b^++1)\pi \Xi \sigma_{ab})^2\Biggr) \ .
\end{align}
Here, we have an exponential decay multiplied with a series in integer powers of $\Xi$, making a different rescaling compared to $P_s(x_s)$ necessary. 
\end{document}